\documentclass[amsmath,amssymb,aps, prb, twocolumn, superscriptaddress]{revtex4-2}

\usepackage{graphicx}

\usepackage{xcolor}

\begin{document}


\title{Abrupt crystallization from shock-compressed CaSiO$_3$ glass}



\author{A. Amouretti}
\email[]{AmourettiA@eie.eng.osaka-u.ac.jp}
\affiliation{Graduate School of Engineering, The University of Osaka, Suita, Osaka 565-0871, Japan}

\author{K. Nonaka}
\affiliation{Graduate School of Engineering, The University of Osaka, Suita, Osaka 565-0871, Japan}

\author{X. Liu}
\affiliation{School of Science, Wuhan University of Technology, Hongshan, Wuhan, Hubei 430074, China}

\author{Y. Hironaka}
\affiliation{Graduate School of Engineering, The University of Osaka, Suita, Osaka 565-0871, Japan}

\author{H. Huang}
\affiliation{School of Science, Wuhan University of Technology, Hongshan, Wuhan, Hubei 430074, China}

\author{R. Kodama}
\affiliation{Graduate School of Engineering, The University of Osaka, Suita, Osaka 565-0871, Japan}
\affiliation{Institute of Laser Engineering, The University of Osaka, Suita, Osaka 565-0871, Japan}

\author{K. Lawler}
\affiliation{Nevada Extreme Conditions Laboratory, University of Nevada, Las Vegas, Nevada 89154, United States}

\author{K. Miyanishi}
\affiliation{RIKEN SPring-8 Center, Hyogo 679-5148, Japan}

\author{H. Nakamura}
\affiliation{Graduate School of Engineering, The University of Osaka, Suita, Osaka 565-0871, Japan}

\author{C. Schwartz}
\affiliation{Nevada Extreme Conditions Laboratory, University of Nevada, Las Vegas, Nevada 89154, United States}

\author{Y. Seto}
\affiliation{Graduate School of Science, Osaka Metropolitan University, Osaka 558-0022, Japan}

\author{K. Sueda}
\affiliation{RIKEN SPring-8 Center, Hyogo 679-5148, Japan}

\author{Y. Wu}
\affiliation{School of Science, Wuhan University of Technology, Hongshan, Wuhan, Hubei 430074, China}

\author{M. Yabashi}
\affiliation{RIKEN SPring-8 Center, Hyogo 679-5148, Japan}
\affiliation{Japan Synchrotron Radiation Research Institute, Hyogo 679-5198, Japan}

\author{T. Yabuuchi}
\affiliation{RIKEN SPring-8 Center, Hyogo 679-5148, Japan}
\affiliation{Japan Synchrotron Radiation Research Institute, Hyogo 679-5198, Japan}

\author{N. Ozaki}
\affiliation{Graduate School of Engineering, The University of Osaka, Suita, Osaka 565-0871, Japan}
\affiliation{Institute of Laser Engineering, The University of Osaka, Suita, Osaka 565-0871, Japan}


\date{\today}

\begin{abstract}
We have performed \textit{in situ} time-resolved X-ray diffraction at $\simeq$100 GPa on laser-shocked CaSiO$_3$ glass to investigate the glass-to-crystal transition. At this extreme pressure, we observe the ultrafast crystallization of the CaSiO$_3$ perovskite structure from the compressed amorphous phase, with a typical nucleation time of 1.69 $\pm$ 0.10~ns and a final grainsize of $\sim$20 nm. The grain size temporal evolution suggest a diffusion controlled transformation.
Moreover, the observed concomitant explosive grain growth together with the release wave arrival into shocked CaSiO$_3$ also suggests a role of the release in the nucleation process. 
\end{abstract}


\maketitle

\section{Introduction} 

The crystal-to-glass or glass-to-crystal transition is a longstanding unsolved problem in modern physics \cite{schmelzer_crystallization_2016,sosso_crystal_2016}. 
The dynamics and kinetics on a short time scale are essential for a profound understanding of transition processes and are critical for further developing the related areas of material science and industry. 
Shock compression experiments are a unique approach to generating such short-duration processes in extreme pressure and temperature conditions. 
Combined laser-shock compression and x-ray diffraction (XRD) diagnostics with x-ray free electron lasers (XFEL) enable visualization of the crystalline structure under the transition phenomena with unprecedented femtosecond time resolution \cite{cerantola_new_2021}. 
Although crystal-to-crystal phase transitions have been studied using this combination \cite{mcbride_phase_2019,pepin_kinetics_2019}, 
the glass transitions have been even less explored. 

Many shock experiments have investigated silicates because of their importance in optics \cite{alexander_changes_2008,carr_localized_2004}, material science, and geoscience \cite{hernandez_direct_2020,okuchi_ultrafast_2021,gleason_ultrafast_2022,kim_femtosecond_2021}. 
Only a few representative XFEL and synchrotron experiments have been conducted on glass silicate, 
and different crystallization kinetics under shock have been observed \cite{gleason_ultrafast_2015,tracy_situ_2018,morard_situ_2020}. 
Fused silica SiO$_2$ crystallization towards stishovite was reported to be surprisingly rapid on the nanosecond timescale \cite{gleason_ultrafast_2015,tracy_situ_2018}. 
In their cases, homogeneous nucleation has been considered to describe processes 
in which the self-diffusion coefficient might play an important roles in crystallization kinetics \cite{shen_nanosecond_2016}. 
So far, the only study on silicate glass including another element was realized on MgSiO$_3$, 
and no observation of crystallization at such a fast timescale has been reported \cite{morard_situ_2020,hernandez_direct_2020}. 
Therefore, performing nanosecond laser shock studies on other complex silicate glasses, {\it e.g.}, 
CaSiO$_3$, is crucial to gaining insight into the silicate glass-to-crystal transition.
Indeed, Ca cations are believed to have higher mobility than Mg in the SiO$_2$ network \cite{bajgain_first-principles_2015}, making CaSiO$_3$ more likely to crystallize than MgSiO$_3$ on a laser shock timescale.

From diamond anvil cells (DAC) static compression and computational studies \cite{liu_synthesis_1975,swamy_thermodynamic_1997,stixrude_phase_2007,noguchi_high-temperature_2013,kawai_pvt_2014,sokolova_equations_2021,milani_crystal_2021,immoor_weak_2022,sun_high-pressure_2022}, 
the CaSiO$_3$ high-pressure phase structure is well-known as a perovskite (Pv-CaSiO$_3$, also named Davemaoite \cite{tschauner_discovery_2021}). 
This phase is stable above 12 GPa and up to 135 GPa at room temperature. 
Metastable CaSiO$_3$ glass can also exist up to $\sim$40 GPa \cite{geballe_sound_2022}. 
While the melting behavior of Pv-CaSiO$_3$ remains debated, it has a calculated high melting temperature of 5000 K or much more at 100 GPa \cite{braithwaite_melting_2019,hernandez_ab_2022,yin_davemaoite_2023}. 
To reach best conditions to observe such a glass-to-crystal transition in CaSiO$_3$ on the nanosecond timescale, high temperatures to promote high nucleation rate in the solid range are required, which might be expected along the Hugoniot near 100 GPa, below the melting curve.
The shock temperature is estimated using the Hugoniot data for CaSiO$_3$ glass obtained recently by some coauthors \cite{wu_hugoniot_2024}, 
and the calculations results detailed later. 

We present here a time-resolved XRD study on the glass-to-crystal transition in CaSiO$_3$-glass laser-shock compressed to 108 GPa.
At this pressure and on this nanosecond timescale, the \textit{in situ} XRD observation reveals the crystallization from the glass into Pv-CaSiO$_3$, in contrast to MgSiO$_3$. We report the time evolution of grain size, associated with the crystallization dynamics and kinetics.

\section{Method} 
\begin{figure}[!h]
\center
\includegraphics[scale=0.55]{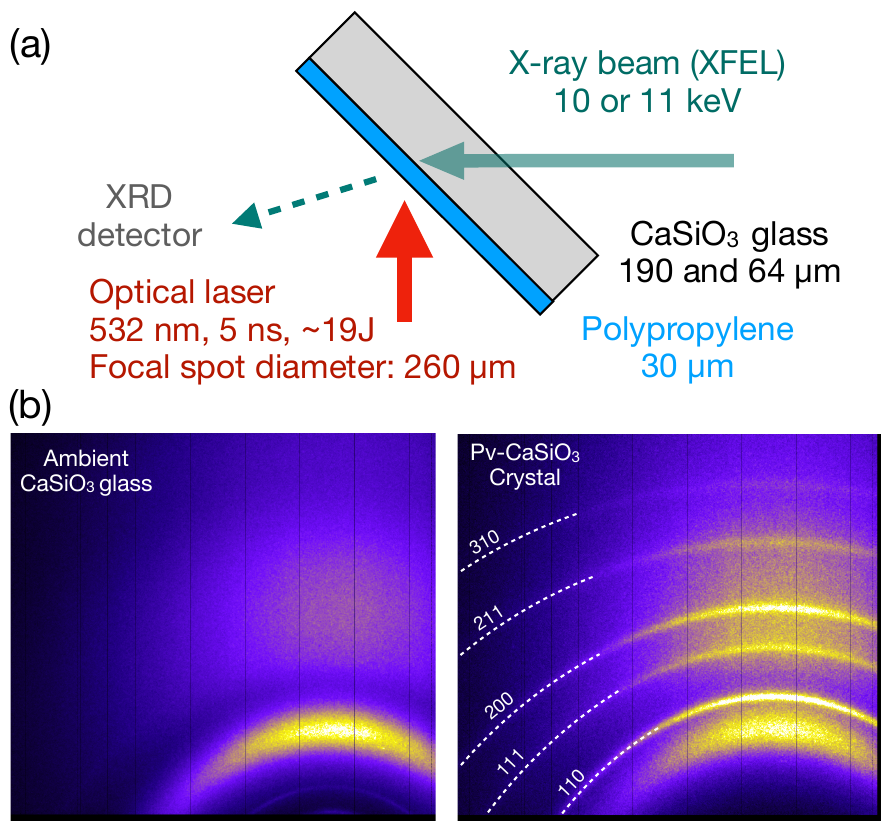}
\caption{\textbf{(a)} Schematic of the experimental configuration and target design.  \textbf{(b)} 2D X-ray diffraction (XRD) patterns signal acquired with the flatpanel detector, showing on left the amorphous signals from the unshocked sample, and on right crystallized Pv-CaSiO$_3$ signals from the shot sample, corresponding to a pump-probe delay of 8.4 ns.
}
\label{setup}
\end{figure}

The laser shock experiments have been performed at the BL3 beamline of the SACLA (SPring-8 Angstrom Compact Free Electron Laser) facility \cite{inubushi_development_2020,inubushi_measurement_2017,ishikawa_compact_2012}. The setup is schematically shown in Fig.\ref{setup}: a 5~ns square laser pulse was focused on the ablator surface of the target with a 260~\textmu m focal spot diameter and a maximum energy of 19~J. The spot intensity distribution was smoothed using a phase plate.
The shock driving laser wavelength was 532 nm.
The XFEL beam arrived at the target rear side with an angle of 45\textdegree~and a focal spot of 10 $\times$ 30~(H $\times$ W)~\textmu m.  The XFEL beam was set to a photon energy of 10 or 11~keV (dE/E $\approx 0.3\%$) and a pulse duration of 10~fs.
The signal diffracted by the target is measured in transmission using a flat panel detector placed about 130~mm behind the target.

Targets consist of a 30~\textmu m thick polypropylene (PP30) ablator, glued by UV-cured adhesive on a 190 $\pm$ 4~\textmu m or 64 $\pm$ 18~\textmu m thick CaSiO$_3$-glass sample. 
The front surface of the polypropylene is made opaque to absorb the laser and avoid shine-through direct irradiation of the CaSiO$_3$ surface due to a possible pre-pulse at the beginning of the laser pulse profile. 
The CaSiO$_3$-glass samples were synthesized at Wuhan University of Technology, with a measured initial density of 2.902(5) g/cm$^3$ \cite{wu_hugoniot_2024}.

Two Velocity Interferometer Systems for Any Reflector (VISAR) \cite{barker_laser_1972} were used to determine indirectly the shock pressure in CaSiO$_3$. As CaSiO$_3$ shock front does not become metallic at 100 GPa, we carried out independent measurements of interface velocities with PP30/LiF targets and measurements of transit time in ablator with PP30 only targets (Fig. S1 of the supplementary).
The VISAR velocity sensitivities were 5.375 and 8.597~km/s/fringes in vacuum, and the sweep time window of streak cameras was 10 ns or 20 ns.
The estimated average shock pressure reached in CaSiO$_3$ during the experiments was 108 +/- 11 GPa. The pump-probe delay between XFEL and shock-driving laser has been tuned to track crystalline change during shock and release. The measured transit time in PP30, is systematically related to the delay and sets the time origin equal to the shock arrival in CaSiO$_3$.

\section{Results}

\begin{figure}[!h]
\center
\includegraphics[scale=0.4]{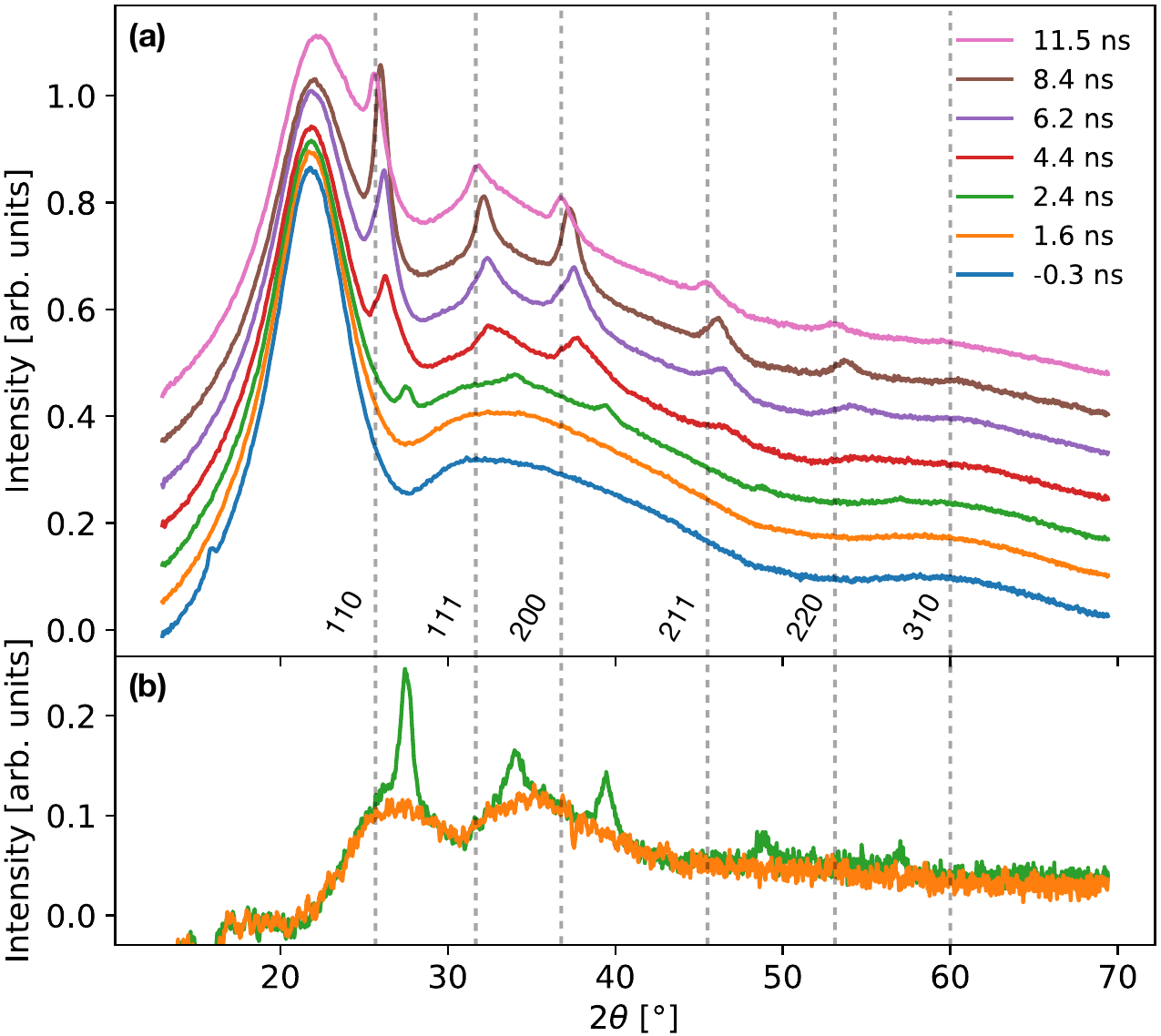}
\caption{
\textbf{(a)} Selection of azimuthally integrated diffraction pattern, obtain with a 11 keV incident x-ray energy, ordered as a function of the arrival time of the shock in CaSiO$_3$. The average pressure for those shots is $108\pm11$~GPa. The vertical dashed lines correspond to the position of the Pv-CaSiO$_3$ ambient peaks. \textbf{(b)} Patterns at 1.8 ns and 2.4 ns with the unshocked ambient signal subtracted, highlighting the presence of a compressed amorphous diffuse signal, characterized by 2 broad peaks around 27$^\circ$ and 35$^\circ$.
}
\label{diff10GPa}
\end{figure}

Fig.\ref{diff10GPa}(a) shows a series of azimuthally integrated diffraction patterns for delays ranging from -0.3 ns to 11.5 ns.
The signal drawn in blue, measured before 0~ns, is used as a reference signal from ambient CaSiO$_3$ glass. At 2.4~ns we observe the appearance of 5 crystalline peaks, not observable at -0.3~ns and 1.6~ns, that can be indexed by (110), (111), (200), (211) and (220) peaks from a high-pressure Pv-CaSiO$_3$ structure, with a unit cell volume of V$_{Pv}$ = 36.94 Å$^3$.
Fig.\ref{diff10GPa}(b) emphasizes the two diffraction patterns at 1.8 ns and 2.4 ns timing, for which the reference diffuse signal from ambient CaSiO$_3$-glass is subtracted. The signal at 1.6 ns shows two broad peaks located around $2\theta = 27^\circ$ (Q = 2.6 Å$^{-1}$) and $35^\circ$ (3.4 Å$^{-1}$), characteristic of a disordered phase, interpreted as the signal from compressed amorphous CaSiO$_3$. A similar high-pressure amorphous signal is observed in other silicates under shock, such as in MgSiO$_3$ \cite{morard_situ_2020,hernandez_direct_2020}. We then notice, at 2.4 ns, mixing of the amorphous signal and crystallization peaks of the high-pressure Pv-CaSiO$_3$ structure, showing the crystallization from the compressed amorphous phase.
Finally, the 2D pattern of crystalline Pv-CaSiO$_3$, as shown in Fig. \ref{setup}(b), highlights extended and non-textured Debye-Scherrer rings. This suggests the absence of a preferred orientation of the Pv-CaSiO$_3$ crystal grains, similarly to studies realized on SiO$_2$ shocked glass \cite{tracy_situ_2018}.

The average grain size of Pv-CaSiO$_3$ crystallites, $d$, is estimated here from the peaks broadening by using the Warren–Averbach \cite{warren_effect_1950} method. 
The full width at half maximum of each of the five indexed peaks of Pv-CaSiO$_3$ listed above is extracted by fitting a pseudo-Voigt function after removing the compressed amorphous signal using a spline function. The results as a function of the pump–probe delay are shown in Fig.~\ref{grain_size}(a), while further details of the method are described in the Supplementary Information (Section III). Grain size values from diffraction patterns with strongly asymmetric peaks, such as at 4.4, 6.2 ns and 11.5 ns, were not plotted, as the broadening is believed to result primarily from significant pressure gradients in the sample rather than from grain size or strain (see the hydrodynamic simulation reported in Fig.~S2 of the SI).


\begin{figure}[!h]
\center
\includegraphics[scale=0.29]{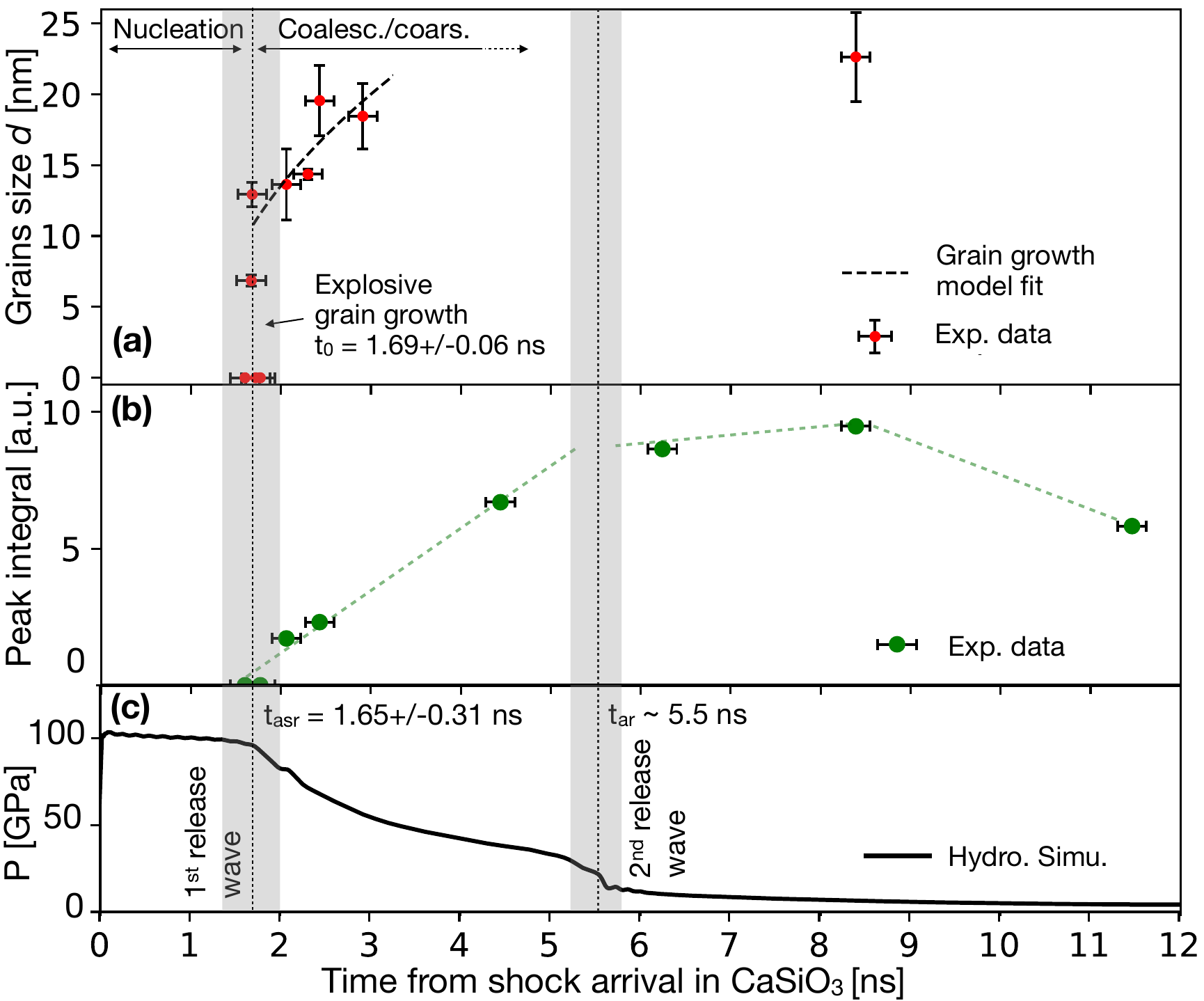}
\caption{\textbf{(a)} Grain size of Pv-CaSiO$_3$ extracted from the diffraction peaks with the Warren-Averbach method \cite{warren_effect_1950}. Best fitting result of formula (\ref{eq_gg}) is given with $n=2$ in the coalescence/coarsening regime and plotted with a dotted black line. \textbf{(b)} Time evolution of the integrated Pv-CaSiO$_3$ peaks intensity. Those values are extracted for all CaSiO$_3$ samples with similar thickness; therefore, they reflect the evolution of the crystal proportion in the sample. \textbf{(c)} Time evolution of pressure at the interface of ablator and LiF obtained through MULTI hydrodynamic simulation (see Fig. S2 in supplementary), calibrated on VISAR results. Two different release wave breakouts are determined from it.}
\label{grain_size}
\end{figure}

In Fig. \ref{grain_size}(a), for the short time range 1.6-1.77 ns, we observe both points without crystallization ($d$~=~0~nm, i.e., only a compressed amorphous phase is observed) and with crystallization ($d>$~0~nm). 
This observation led us to consider an abrupt grain growth, from zero to 14~nm within $\simeq$200~ps, at the timing $t_0=1.69~\pm$~0.10~ns.
At later delays, a progressive increase of the grain size is observed from 14 nm to 20 nm and up to 3 ns. 
The data point at 8.4 ns also shows a grain size of 22 nm, indicating that the grain almost stopped growing around 3 ns. 
As seen in Fig. \ref{grain_size}(b), the relative proportion of Pv-CaSiO$_3$ crystal in the sample, deduced from the peak intensity integration, tends to linearly increase in the sample until ~5.5 ns, and goes also toward a crystallization start around 1.6-1.7~ns. The plateau starting at $\sim$5.5 ns followed by a decrease at 11.4 ns may indicate amorphization or melting at longer times, occurring on very fast timescales, due to important release (SI Section II).

In the previous publications \cite{shen_nanosecond_2016,gleason_ultrafast_2015}, this grain size evolution is interpreted through 3 successive stages: nucleation, explosive grain growth, and coalescence. 
The nucleation stage corresponds to the region before the grain growth jump in our data, when the formed nuclei of Pv-CaSiO$_3$ are too small and few to be observed in the diffraction pattern. 
In the explosive grain growth stage, assigned here to the grain size jump at $\simeq$1.7 ns, enough nuclei reach a critical radius, thus the crystallized CaSiO$_3$ volume increases rapidly with grain size.
After this stage, a coalescence/coarsening stage starts, explaining the slower growth from 1.7 to 3 ns in our data.


The grain size in those different grain growth stages can be modeled by the general formula : 
\begin{equation}
d^n-d_0^n = K(t-t_0)
\label{eq_gg}
\end{equation}
 with $t_0$ the starting time of the process, $n$ an exponent related to the growth mechanism \cite{christian_theory_2002}, $K$ a coefficient that depends on the temperature and the $d_0$ the diameter size of grain at $t_0$. Different values for those parameters are expected for the different stages \cite{shen_nanosecond_2016}.
The explosive grain growth stage is too fast to be fitted, but enough data points in the coalescence/coarsening allows us to use the formula for the range from $1.65$~ns to $t=3$~ns, excluding data with $d=0$~nm. For values of $n$ from 1 to 7, the data of $d^n$ are fitted with linear function, and the values of $K$ and $d_0$ are determined from the linear coefficients (See SI, section III and Fig. S3). We find that a value of $n = 2\pm1$ gave best precision on $d_0$ and $K$.
In Fig. \ref{grain_size}(a) $n=2$ is used to represent the results; however our result also support $n=3$, based on the coarsening law reported in \cite{shen_nanosecond_2016}, as an acceptable value. For $n=2$, we have $d_0 = 10.8\pm2.1$~nm and $K = (2.2\pm0.7)\cdot 10^4$~nm$^3$/ns.
While a similar nucleation time, 1.6 ns, and final grain size, 20 nm, was found in the previous study on SiO$_2$ \cite{gleason_ultrafast_2015}, a value of $n=$7 was used. That might be explained by a different crystallization mechanism due to the higher pressure achieved in our case. The coalescence/coarsening stage in our study seems better modeled by $n\leq4$, suggesting an overall diffusion-controlled transformation \cite{huang_two-stage_2003, lifshitz_kinetics_1961}.


In the Hugoniot measured by Wu \textit{et  al.} \cite{wu_hugoniot_2024}, a linear $u_s$-$u_p$ relation: $u_s=2.95(11)+1.67(5)u_p$ is observed in the $u_p$ range between 1.8 km/s and 3.0 km/s, where $u_s$ is the shock velocity and $u_p$ the particle velocity. 
As CaSiO$_3$ is opaque under shock for pressures higher than 80 GPa \cite{wu_hugoniot_2024} and metallization occurs only at pressures far above 100~GPa, VISAR measurements of shock velocity or particle velocity in CaSiO$_3$ under our conditions was impossible.
Because the LiF Hugoniot is quasi-equal to the extrapolated CaSiO$_3$ glass Hugoniot to 100 GPa (see Fig. \ref{grain_size}(c) and Fig. S2 in SI), we rely on the interface velocity measurements from a target with LiF window to simulate the pressure and shock velocity evolution in CaSiO$_3$ with time, for the same laser energy. Especially, we assumed that the reflected shock waves' velocities in the ablator are the same in both the LiF and CaSiO$_3$ cases.The VISAR measurements and MULTI hydrodynamic simulations \cite{ramis_multi_1988}, evidenced two release waves entering the LiF, after the shock wave arrival in LiF at $t=0$~ns: the first release wave, the ablation surface-release wave \cite{swift_properties_2008}, at $t_{asr} = 1.65 \pm 0.31$~ns, characterized by a gradual pressure decrease in the  PP/LiF interface. The second release wave, the ablation release wave \cite{swift_properties_2008}, at $t_{ar}\sim5.5$~ns, originated from the end of the laser pulse, rapidly decreasing the LiF pressure close to zero.
Those two release-wave breakout times are shown in Fig. \ref{grain_size} by dotted vertical lines. 
We then observe that the first release wave breakout ($t_{asr}$) is concurrent, within uncertainty, with the explosive crystal grain growth ($t_0$), suggesting a correlation between release and crystallization process.

\section{Discussion}
\begin{figure}[!h]
\center
\includegraphics[scale=0.5]{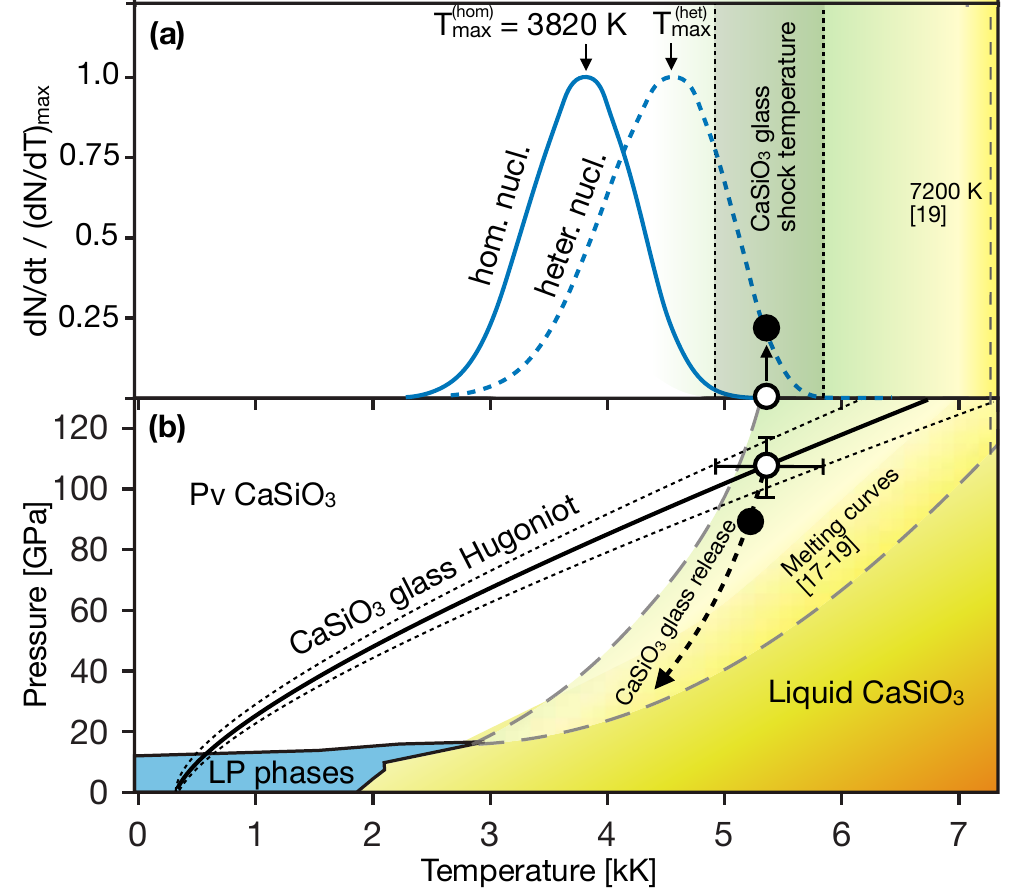}
\caption{\textbf{(a)} Estimation of nucleation rate dependence with temperature, calculated from Eq.~(\ref{nucleation}) for CaSiO$_3$; the curve is calculated using a 7200~K melting temperature at 108 GPa for CaSiO$_3$ given by the melting curve of \cite{yin_davemaoite_2023}. The qualitative representation of nucleation rate for heterogeneous nucleation is shown in blue dotted line. The calculated Hugoniot temperature condition of CaSiO$_3$ glass at 108 GPa is by the grey vertical region. \textbf{(b)} Phase diagram of CaSiO$_3$ with calculated Hugoniot of CaSiO$_3$ glass, and schematic of release path from CaSiO$_3$ glass compressed at 108 GPa. Melting region boundaries are determined from three different melting curves \cite{braithwaite_melting_2019,hernandez_ab_2022,yin_davemaoite_2023}.}
\label{diag}
\end{figure}

The work of Wu \textit{et al.} \cite{wu_hugoniot_2024} shows that, by the absence of discontinuity in sound velocity measurement along the Hugoniot, the shock melting of CaSiO$_3$ glass occurs at a pressure above 116 GPa. Thus, the disordered phase observed at 108 GPa in our study is most likely a metastable amorphous solid.
The Hugoniot temperature of amorphous CaSiO$_3$ is deduced from the Mie-Gruneisen relation \cite{rice_compression_1958}:
\begin{equation}
T(\eta) = T_0e^{\gamma_0\eta} - \frac{V_0e^{\gamma_0\eta}}{C_v}\int_0^{\eta}\biggr[\frac{P(\eta^{\prime})}{2}-\frac{dP}{d\eta^{\prime}}\frac{\eta^{\prime}}{2}\biggl]e^{-\gamma_0\eta'}d\eta'
\label{hug_temp}
\end{equation}
Where $\eta=1-v/v_0$ is the compression, $v$ the specific volume of CaSiO$_3$ glass  and $v_0 = 1/\rho_0 = 0.345~$cm$^{3}\cdot$g$^{\text{-}1}$ its initial value. $P(\eta)$ is the Hugoniot pressure of CaSiO$_3$ glass, evaluated from the Hugoniot linear relation described before \cite{wu_hugoniot_2024}.
The specific heat capacity $C_v=~1039$~J/kg/K and Grüneisen parameter $\gamma_0 = 2$ from the solid Pv-CaSiO$_3$ crystalline phase has been used \cite{bajgain_first-principles_2015}.
~The calculated Hugoniot temperature as a function of pressure is shown in the black line on the phase diagram of Fig.~\ref{diag}(b). The dotted line represent the uncertainty on the temperature, determined from the maximum and minimum values of the Hugoniot coefficients, the Gruneisen parameter, and the heat capacity  \cite{kawai_pvt_2014, bajgain_first-principles_2015,wu_hugoniot_2024}.
For a shock pressure of 108~GPa, the estimated temperature is 5400~$\pm$~500~K.
The Pv-CaSiO$_3$ isothermal equation of states \cite{bajgain_first-principles_2015,sokolova_equations_2021} and the CaSiO$_3$ averaged unit cell volume extract from our XRD measurments at 108~GPa can be used to estimate the Pv-CaSiO$_3$ temperature. Indeed, V$_{Pv}$ = 37.3$\pm$0.7 Å$^3$ at 108 GPa, and the unit cell volume of the Pv-CaSiO$_3$ phase statically compressed at 108 GPa and 300~K is V$_{Pv}(300~\text{K})$ = 35.03 Å$^3$. The thermal expansion coefficient at 108 GPa ($1.21\times10^{-5}$ K$^{-1}$ \cite{bajgain_first-principles_2015}) gives an estimated temperature of Pv-CaSiO$_3$ around $5345\pm1600$~K.
Generally, crystallization is an exothermic process; therefore, the shock temperature of CaSiO$_3$ glass should be lower than that of crystallized Pv-CaSiO$_3$, which gives good confidence in the estimation through relation (\ref{hug_temp}).

As shown in the Results section, the temporal evolution of the grain size indicates a diffusion-controlled transformation during the transition from compressed CaSiO$_3$ glass to Pv-CaSiO$_3$.
In the case of homogeneous nucleation dominated by a diffusion-controlled process, the temperature dependence of the nucleation rate can be described by the Volmer–Weber relation~\cite{callister_fundamentals_2022}; 
\begin{equation}
\frac{dN}{dt} = A\bigg[\exp\bigg(-\alpha\frac{T_m^2}{T(T-T_m)^2}\bigg)\exp\bigg(-\frac{Q_n}{T}\bigg)\bigg]
\label{nucleation}
\end{equation} where $A$ is a constant in ns·nm$^{\text{-}3}$, $T_m$ is the melting temperature, $\alpha$ is a constant depending on the latent heat of fusion and the surface free energy, and $Q_n$ is the hopping parameter associated with the transformation. To predict the temperature corresponding to the maximum nucleation rate using relation (\ref{nucleation}), the parameters were estimated for CaSiO$_3$ silicate glass at 108~GPa, following the reasoning of references \cite{schmelzer_crystallization_2015, weinberg_location_1986} and using the available data presented in articles \cite{yin_davemaoite_2023,bajgain_first-principles_2015,wu_hugoniot_2024}. We estimate $\alpha = 6.3\times10^3$~K and $Q_n = 3.7$~eV (more details are provided in Section IV of the SI). While the high-pressure melting temperature $T_m$ shows a discrepancy of more than 2000 K between different \textit{ab initio} calculations \cite{hernandez_ab_2022,braithwaite_melting_2019,yin_davemaoite_2023}, the highest reported value, $T_m = 7200$~K at 108 GPa from \cite{yin_davemaoite_2023}, was used to set an upper bound for nucleation rate curve. As shown in blue in Fig.~\ref{diag}(a), the variation of $dN/dt/(dN/dt)_{max}$ as a function of temperature $T$ reaches a maximum at $\text{T}^{\text{(hom)}}_{\text{max}} = 3820$~K and remains significant in the approximate temperature range from 3000~K to 4700~K. Thus, the homogeneous nucleation model predicts a temperature for the maximum nucleation rate that is too low compared to the temperature reached under shock (grey error-bar region in Fig.~\ref{diag}(a)) to explain the rapid crystallization observed in our study. 

The simultaneity between the onset of CaSiO$_3$ crystallization and the arrival of the release wave observed in this study also suggests that homogeneous nucleation alone may not fully explain the process. 
If heterogeneous nucleation is triggered by the release wave, it can be viewed as the introduction of an interface at the release front, leading to faster kinetics through a reduced activation energy barrier, $\Delta G^*_{het} = \Delta G^*_{hom} S$ with $0 < S < 1$ \cite{callister_fundamentals_2022}, where $\Delta G^*_{hom} = k_B \alpha T_m^2/(T - T_m)^2$. Generally, the temperature of maximum heterogeneous nucleation, $\mathrm{T}^{(\text{het})}_{\text{max}}$, is shifted toward higher values compared to homogeneous nucleation \cite{callister_fundamentals_2022,schmelzer_crystallization_2015}, which in this case may allow an overlap between the nucleation rate curve and the shock conditions, as illustrated in Fig.~\ref{diag}(a). Pressure gradients at the release front may therefore promote heterogeneous pathways. While this interpretation remains qualitative given our data, it calls for further experimental or molecular dynamics studies.

\section*{Conclusions} 
In conclusion, we present the first evidence of the nanosecond crystallization from CaSiO$_3$ silicate glass to crystal Pv-CaSiO$_3$ under laser shock compression. 
This is the first study on laser shock compressed glass silicate for which a very fast nanosecond crystallization is observed at such high pressure and temperature of 108~GPa and $\simeq$5000~K.
The analysis of grain growth temporal evolution suggest that the transformation from compressed CaSiO$_3$ glass to Pv-CaSiO$_3$ in this condition is a process dominated by diffusion.
Meanwhile, our time-resolved observations suggest that the crystallization process can be divided into three stages: the single-shock Hugoniot, a weakly pressure-releasing quasi-Hugoniot, and the pressure-released state. The abrupt grain growth occurs immediately upon arrival of the release wave at the shocked glass, followed by slower grain coarsening during the subsequent pressure-released state.
Our work encourages new experiments to precisely observe the time evolution in other silicates, even like SiO$_2$ and MgSiO$_3$, for a deeper understanding of the glass-crystal transition. 
Indeed, the crystallization behavior of these silicates has also implications for understanding the history of pressure-induced amorphization 
and subsequent crystallization during asteroid and Martian impacts \cite{tomioka_shockinduced_2000, tomioka_breakdown_2003,xie_host_2007}.

\section*{Acknowledgements} 

This experiment was realized at BL3 of SACLA with the approval of the Japan Synchrotron Radiation Research Institute (proposals No. 2020A8047, No. 2021B8067,  No. 2022A8052, and No. 2022B8044). We thank all the support team and the SACLA staff for their great contribution to this work. The high-power nanosecond laser at SACLA was installed by the Institute of Laser Engineering of Osaka University with the corporation of Hamamatsu Photonics.
This work was supported by grants from Japan Society for the Promotion of Science (JSPS) KAKENHI (Grant No. 22KF0243, No. 20H00139, No. 22K18702) and Core-to-Core Program (No. JPJSCCA20230003), and MEXT Quantum Leap Flagship Program (Grant No. JPMXS0118067246). This work was supported by grants from the National Natural Science Foundation of China (No. 42274123). CPS and KVL were supported by the U.S. Deparment of Energy under grant DE-SC0023355 program for clean energy manufacturing.
\clearpage

\bibliographystyle{apsrev4-2.bst}
\bibliography{CaSiO3_paper}

@article{sokolova_equations_2021,
	title = {Equations of {State} of {Ca}-{Silicates} and {Phase} {Diagram} of the {CaSiO3} {System} under {Upper} {Mantle} {Conditions}},
	volume = {11},
	issn = {2075-163X},
	url = {https://www.mdpi.com/2075-163X/11/3/322},
	doi = {10.3390/min11030322},
	 
	number = {3},
	urldate = {2022-07-20},
	journal = {Minerals},
	author = {Sokolova, Tatiana S. and Dorogokupets, Peter I.},
	month = mar,
	year = {2021},
	pages = {322},
}

@article{milani_crystal_2021,
	title = {Crystal {Structure} {Evolution} of {CaSiO3} {Polymorphs} at {Earth}’s {Mantle} {Pressures}},
	volume = {11},
	issn = {2075-163X},
	url = {https://www.mdpi.com/2075-163X/11/6/652},
	doi = {10.3390/min11060652},
	 
	number = {6},
	urldate = {2022-07-20},
	journal = {Minerals},
	author = {Milani, Sula and Comboni, Davide and Lotti, Paolo and Fumagalli, Patrizia and Ziberna, Luca and Maurice, Juliette and Hanfland, Michael and Merlini, Marco},
	month = jun,
	year = {2021},
	pages = {652},
}

@article{swift_properties_2008,
	title = {On the {Properties} of {Plastic} {Ablators} in {Laser}-{Driven} {Material} {Dynamics} {Experiments}},
	volume = {77},
	issn = {1539-3755, 1550-2376},
	url = {http://arxiv.org/abs/0712.1203},
	doi = {10.1103/PhysRevE.77.066402},
	number = {6},
	urldate = {2022-04-07},
	journal = {Phys. Rev. E},
	author = {Swift, Damian C. and Kraus, Richard G.},
	month = jun,
	year = {2008},
	note = {arXiv: 0712.1203},
	pages = {066402},
}

@article{mcbride_phase_2019,
	title = {Phase transition lowering in dynamically compressed silicon},
	volume = {15},
	issn = {1745-2473, 1745-2481},
	url = {http://www.nature.com/articles/s41567-018-0290-x},
	doi = {10.1038/s41567-018-0290-x},
	 
	number = {1},
	urldate = {2021-11-21},
	journal = {Nature Phys},
	author = {McBride, E. E. and Krygier, A. and Ehnes, A. and Galtier, E. and Harmand, M. and Konôpková, Z. and Lee, H. J. and Liermann, H.-P. and Nagler, B. and Pelka, A. and Rödel, M. and Schropp, A. and Smith, R. F. and Spindloe, C. and Swift, D. and Tavella, F. and Toleikis, S. and Tschentscher, T. and Wark, J. S. and Higginbotham, A.},
	month = jan,
	year = {2019},
	pages = {89--94},
}

@article{kim_femtosecond_2021,
	title = {Femtosecond {X}‐{Ray} {Diffraction} of {Laser}‐{Shocked} {Forsterite} ({Mg2SiO4}) to 122 {GPa}},
	volume = {126},
	issn = {2169-9313, 2169-9356},
	url = {https://onlinelibrary.wiley.com/doi/10.1029/2020JB020337},
	doi = {10.1029/2020JB020337},
	 
	number = {1},
	urldate = {2021-11-19},
	journal = {J Geophys Res Solid Earth},
	author = {Kim, Donghoon and Tracy, Sally J. and Smith, Raymond F. and Gleason, Arianna E. and Bolme, Cindy A. and Prakapenka, Vitali B. and Appel, Karen and Speziale, Sergio and Wicks, June K. and Berryman, Eleanor J. and Han, Sirus K. and Schoelmerich, Markus O. and Lee, Hae Ja and Nagler, Bob and Cunningham, Eric F. and Akin, Minta C. and Asimow, Paul D. and Eggert, Jon H. and Duffy, Thomas S.},
	month = jan,
	year = {2021},
}

@article{cerantola_new_2021,
	title = {New frontiers in extreme conditions science at synchrotrons and free electron lasers},
	volume = {33},
	issn = {0953-8984, 1361-648X},
	url = {https://iopscience.iop.org/article/10.1088/1361-648X/abfd50},
	doi = {10.1088/1361-648X/abfd50},
	number = {27},
	urldate = {2021-07-08},
	journal = {J. Phys.: Condens. Matter},
	author = {Cerantola, Valerio and Rosa, Angelika Dorothea and Konôpková, Zuzana and Torchio, Raffaella and Brambrink, Erik and Rack, Alexander and Zastrau, Ulf and Pascarelli, Sakura},
	month = jul,
	year = {2021},
	pages = {274003},
}

@article{hernandez_direct_2020,
	title = {Direct {Observation} of {Shock}‐{Induced} {Disordering} of {Enstatite} {Below} the {Melting} {Temperature}},
	volume = {47},
	issn = {0094-8276, 1944-8007},
	url = {https://onlinelibrary.wiley.com/doi/10.1029/2020GL088887},
	doi = {10.1029/2020GL088887},
	 
	number = {15},
	urldate = {2021-07-08},
	journal = {Geophys. Res. Lett.},
	author = {Hernandez, J.‐A. and Morard, G. and Guarguaglini, M. and Alonso‐Mori, R. and Benuzzi‐Mounaix, A. and Bolis, R. and Fiquet, G. and Galtier, E. and Gleason, A. E. and Glenzer, S. and Guyot, F. and Ko, B. and Lee, H. J. and Mao, W. L. and Nagler, B. and Ozaki, N. and Schuster, A. K. and Shim, S. H. and Vinci, T. and Ravasio, A.},
	month = aug,
	year = {2020},
}

@article{shen_nanosecond_2016,
	title = {Nanosecond homogeneous nucleation and crystal growth in shock-compressed {SiO2}},
	volume = {15},
	issn = {1476-1122, 1476-4660},
	url = {http://www.nature.com/articles/nmat4447},
	doi = {10.1038/nmat4447},
	 
	number = {1},
	urldate = {2021-07-08},
	journal = {Nature Mater},
	author = {Shen, Yuan and Jester, Shai B. and Qi, Tingting and Reed, Evan J.},
	month = jan,
	year = {2016},
	pages = {60--65},
}

@article{gleason_ultrafast_2015,
	title = {Ultrafast visualization of crystallization and grain growth in shock-compressed {SiO2}},
	volume = {6},
	issn = {2041-1723},
	url = {http://www.nature.com/articles/ncomms9191},
	doi = {10.1038/ncomms9191},
	 
	number = {1},
	urldate = {2021-07-08},
	journal = {Nat Commun},
	author = {Gleason, A. E. and Bolme, C. A. and Lee, H. J. and Nagler, B. and Galtier, E. and Milathianaki, D. and Hawreliak, J. and Kraus, R. G. and Eggert, J. H. and Fratanduono, D. E. and Collins, G. W. and Sandberg, R. and Yang, W. and Mao, W. L.},
	month = nov,
	year = {2015},
	pages = {8191},
}

@article{morard_situ_2020,
	title = {In situ {X}-ray diffraction of silicate liquids and glasses under dynamic and static compression to megabar pressures},
	volume = {117},
	issn = {0027-8424, 1091-6490},
	url = {http://www.pnas.org/lookup/doi/10.1073/pnas.1920470117},
	doi = {10.1073/pnas.1920470117},
	 
	number = {22},
	urldate = {2021-05-14},
	journal = {Proc Natl Acad Sci USA},
	author = {Morard, Guillaume and Hernandez, Jean-Alexis and Guarguaglini, Marco and Bolis, Riccardo and Benuzzi-Mounaix, Alessandra and Vinci, Tommaso and Fiquet, Guillaume and Baron, Marzena A. and Shim, Sang Heon and Ko, Byeongkwan and Gleason, Arianna E. and Mao, Wendy L. and Alonso-Mori, Roberto and Lee, Hae Ja and Nagler, Bob and Galtier, Eric and Sokaras, Dimosthenis and Glenzer, Siegfried H. and Andrault, Denis and Garbarino, Gaston and Mezouar, Mohamed and Schuster, Anja K. and Ravasio, Alessandra},
	month = jun,
	year = {2020},
	pages = {11981--11986},
}

@article{barker_laser_1972,
	title = {Laser interferometer for measuring high velocities of any reflecting surface},
	volume = {43},
	issn = {0021-8979, 1089-7550},
	url = {http://aip.scitation.org/doi/10.1063/1.1660986},
	doi = {10.1063/1.1660986},
	 
	number = {11},
	urldate = {2020-07-30},
	journal = {Journal of Applied Physics},
	author = {Barker, L. M. and Hollenbach, R. E.},
	month = nov,
	year = {1972},
	pages = {4669--4675},
}

@article{inubushi_development_2020,
	title = {Development of an {Experimental} {Platform} for {Combinative} {Use} of an {XFEL} and a {High}-{Power} {Nanosecond} {Laser}},
	volume = {10},
	issn = {2076-3417},
	url = {https://www.mdpi.com/2076-3417/10/7/2224},
	doi = {10.3390/app10072224},
	 
	number = {7},
	urldate = {2020-07-17},
	journal = {Applied Sciences},
	author = {Inubushi, Yuichi and Yabuuchi, Toshinori and Togashi, Tadashi and Sueda, Keiichi and Miyanishi, Kohei and Tange, Yoshinori and Ozaki, Norimasa and Matsuoka, Takeshi and Kodama, Ryosuke and Osaka, Taito and Matsuyama, Satoshi and Yamauchi, Kazuto and Yumoto, Hirokatsu and Koyama, Takahisa and Ohashi, Haruhiko and Tono, Kensuke and Yabashi, Makina},
	month = mar,
	year = {2020},
	pages = {2224},
}

@article{pepin_kinetics_2019,
	title = {Kinetics and structural changes in dynamically compressed bismuth},
	volume = {100},
	issn = {2469-9950, 2469-9969},
	url = {https://link.aps.org/doi/10.1103/PhysRevB.100.060101},
	doi = {10.1103/PhysRevB.100.060101},
	 
	number = {6},
	urldate = {2020-07-13},
	journal = {Phys. Rev. B},
	author = {Pépin, Charles M. and Sollier, Arnaud and Marizy, Adrien and Occelli, Florent and Sander, Mathias and Torchio, Raffaella and Loubeyre, Paul},
	month = aug,
	year = {2019},
	pages = {060101},
}

@article{tracy_situ_2018,
	title = {\textit{{In} situ} {X}-{Ray} {Diffraction} of {Shock}-{Compressed} {Fused} {Silica}},
	volume = {120},
	issn = {0031-9007, 1079-7114},
	url = {https://link.aps.org/doi/10.1103/PhysRevLett.120.135702},
	doi = {10.1103/PhysRevLett.120.135702},
	 
	number = {13},
	urldate = {2022-10-20},
	journal = {Phys. Rev. Lett.},
	author = {Tracy, Sally June and Turneaure, Stefan J. and Duffy, Thomas S.},
	month = mar,
	year = {2018},
	pages = {135702},
}

@article{braithwaite_melting_2019,
	title = {Melting of {CaSiO} $_{\textrm{3}}$ {Perovskite} at {High} {Pressure}},
	volume = {46},
	issn = {0094-8276, 1944-8007},
	url = {https://onlinelibrary.wiley.com/doi/abs/10.1029/2018GL081805},
	doi = {10.1029/2018GL081805},
	 
	number = {4},
	urldate = {2022-11-16},
	journal = {Geophys. Res. Lett.},
	author = {Braithwaite, James and Stixrude, Lars},
	month = feb,
	year = {2019},
	pages = {2037--2044},
}

@article{liu_synthesis_1975,
	title = {Synthesis of a perovskite-type polymorph of {CaSiO3}},
	volume = {28},
	issn = {0012821X},
	url = {https://linkinghub.elsevier.com/retrieve/pii/0012821X75902290},
	doi = {10.1016/0012-821X(75)90229-0},
	 
	number = {2},
	urldate = {2022-11-16},
	journal = {Earth and Planetary Science Letters},
	author = {Liu, Lin-Gun and Ringwood, A.E.},
	month = dec,
	year = {1975},
	pages = {209--211},
}

@article{geballe_sound_2022,
	title = {Sound speed and refractive index of amorphous {CaSiO3} upon pressure cycling to 40 {GPa}},
	volume = {107},
	issn = {0003-004X, 1945-3027},
	url = {https://pubs.geoscienceworld.org/ammin/article/107/12/2212/619049/Sound-speed-and-refractive-index-of-amorphous},
	doi = {10.2138/am-2022-8081},
	 
	number = {12},
	urldate = {2022-12-13},
	journal = {American Mineralogist},
	author = {Geballe, Zachary M. and Arveson, Sarah M. and Speziale, Sergio and Jeanloz, Raymond},
	month = dec,
	year = {2022},
	pages = {2212--2218},
}

@article{swamy_thermodynamic_1997,
	title = {Thermodynamic data for the phases in the {CaSiO3} system},
	volume = {61},
	issn = {00167037},
	url = {https://linkinghub.elsevier.com/retrieve/pii/S0016703796004036},
	doi = {10.1016/S0016-7037(96)00403-6},
	 
	number = {6},
	urldate = {2022-12-16},
	journal = {Geochimica et Cosmochimica Acta},
	author = {Swamy, Varghese and Dubrovinsky, Leonid S.},
	month = mar,
	year = {1997},
	pages = {1181--1191},
}

@article{bajgain_first-principles_2015,
	title = {First-principles simulations of {CaO} and {CaSiO3} liquids: structure, thermodynamics and diffusion},
	volume = {42},
	issn = {0342-1791, 1432-2021},
	shorttitle = {First-principles simulations of {CaO} and {CaSiO3} liquids},
	url = {http://link.springer.com/10.1007/s00269-014-0730-9},
	doi = {10.1007/s00269-014-0730-9},
	 
	number = {5},
	urldate = {2022-12-16},
	journal = {Phys Chem Minerals},
	author = {Bajgain, Suraj K. and Ghosh, Dipta B. and Karki, Bijaya B.},
	month = may,
	year = {2015},
	pages = {393--404},
}

@article{gleason_ultrafast_2022,
	title = {Ultrafast structural response of shock‐compressed plagioclase},
	volume = {57},
	issn = {1086-9379, 1945-5100},
	url = {https://onlinelibrary.wiley.com/doi/10.1111/maps.13785},
	doi = {10.1111/maps.13785},
	 
	number = {3},
	urldate = {2023-06-08},
	journal = {Meteorit \& Planetary Scien},
	author = {Gleason, Arianna E. and Park, Sulgiye and Rittman, Dylan R. and Ravasio, Alessandra and Langenhorst, Falko and Bolis, Riccardo M. and Granados, Eduardo and Hok, Sovanndara and Kroll, Thomas and Sikorski, Marcin and Weng, Tsu‐Chien and Lee, Hae Ja and Nagler, Bob and Sisson, Thomas and Xing, Zhou and Zhu, Diling and Giuli, Gabriele and Mao, Wendy L. and Glenzer, Siegfried H. and Sokaras, Dimosthenis and Alonso‐Mori, Roberto},
	month = mar,
	year = {2022},
	pages = {635--643},
}

@article{tschauner_discovery_2021,
	title = {Discovery of davemaoite, {CaSiO3}-perovskite, as a mineral from the lower mantle},
	volume = {374},
	issn = {0036-8075, 1095-9203},
	url = {https://www.science.org/doi/10.1126/science.abl8568},
	doi = {10.1126/science.abl8568},
	 
	number = {6569},
	urldate = {2023-06-27},
	journal = {Science},
	author = {Tschauner, Oliver and Huang, Shichun and Yang, Shuying and Humayun, Munir and Liu, Wenjun and Gilbert Corder, Stephanie N and Bechtel, Hans A. and Tischler, Jon and Rossman, George R.},
	month = nov,
	year = {2021},
	pages = {891--894},
}

@article{hernandez_ab_2022,
	title = {Ab {Initio} {Atomistic} {Simulations} of {Ca}‐{Perovskite} {Melting}},
	volume = {49},
	issn = {0094-8276, 1944-8007},
	url = {https://onlinelibrary.wiley.com/doi/10.1029/2021GL097262},
	doi = {10.1029/2021GL097262},
	 
	number = {20},
	urldate = {2023-07-25},
	journal = {Geophysical Research Letters},
	author = {Hernandez, J.‐A. and Mohn, C. E. and Guren, M. G. and Baron, M. A. and Trønnes, R. G.},
	month = oct,
	year = {2022},
}

@article{tomioka_breakdown_2003,
	title = {The breakdown of diopside to {Ca}-rich majorite and glass in a shocked {H} chondrite},
	volume = {208},
	issn = {0012821X},
	url = {https://linkinghub.elsevier.com/retrieve/pii/S0012821X03000499},
	doi = {10.1016/S0012-821X(03)00049-9},
	 
	number = {3-4},
	urldate = {2023-07-28},
	journal = {Earth and Planetary Science Letters},
	author = {Tomioka, Naotaka and Kimura, Makoto},
	month = mar,
	year = {2003},
	pages = {271--278},
}

@article{sosso_crystal_2016,
	title = {Crystal {Nucleation} in {Liquids}: {Open} {Questions} and {Future} {Challenges} in {Molecular} {Dynamics} {Simulations}},
	volume = {116},
	issn = {0009-2665, 1520-6890},
	shorttitle = {Crystal {Nucleation} in {Liquids}},
	url = {https://pubs.acs.org/doi/10.1021/acs.chemrev.5b00744},
	doi = {10.1021/acs.chemrev.5b00744},
	 
	number = {12},
	urldate = {2023-09-28},
	journal = {Chem. Rev.},
	author = {Sosso, Gabriele C. and Chen, Ji and Cox, Stephen J. and Fitzner, Martin and Pedevilla, Philipp and Zen, Andrea and Michaelides, Angelos},
	month = jun,
	year = {2016},
	pages = {7078--7116},
}

@article{yin_davemaoite_2023,
	title = {Davemaoite as the mantle mineral with the highest melting temperature},
	volume = {9},
	issn = {2375-2548},
	url = {https://www.science.org/doi/10.1126/sciadv.adj2660},
	doi = {10.1126/sciadv.adj2660},
	 
	number = {49},
	urldate = {2023-12-19},
	journal = {Sci. Adv.},
	author = {Yin, Kun and Belonoshko, Anatoly B. and Li, Yonghui and Lu, Xiancai},
	month = dec,
	year = {2023},
	pages = {eadj2660},
}

@article{xie_host_2007,
	title = {Host rock solid-state transformation in a shock-induced melt vein of {Tenham} {L6} chondrite},
	volume = {254},
	issn = {0012821X},
	url = {https://linkinghub.elsevier.com/retrieve/pii/S0012821X06008685},
	doi = {10.1016/j.epsl.2006.12.001},
	 
	number = {3-4},
	urldate = {2024-02-09},
	journal = {Earth and Planetary Science Letters},
	author = {Xie, Zhidong and Sharp, Thomas G.},
	month = feb,
	year = {2007},
	pages = {433--445},
}

@article{inubushi_measurement_2017,
	title = {Measurement of the {X}-ray {Spectrum} of a {Free} {Electron} {Laser} with a {Wide}-{Range} {High}-{Resolution} {Single}-{Shot} {Spectrometer}},
	volume = {7},
	issn = {2076-3417},
	url = {http://www.mdpi.com/2076-3417/7/6/584},
	doi = {10.3390/app7060584},
	 
	number = {6},
	urldate = {2024-02-22},
	journal = {Applied Sciences},
	author = {Inubushi, Yuichi and Inoue, Ichiro and Kim, Jangwoo and Nishihara, Akihiko and Matsuyama, Satoshi and Yumoto, Hirokatsu and Koyama, Takahisa and Tono, Kensuke and Ohashi, Haruhiko and Yamauchi, Kazuto and Yabashi, Makina},
	month = jun,
	year = {2017},
	pages = {584},
}

@article{ishikawa_compact_2012,
	title = {A compact {X}-ray free-electron laser emitting in the sub-ångström region},
	volume = {6},
	issn = {1749-4885, 1749-4893},
	url = {https://www.nature.com/articles/nphoton.2012.141},
	doi = {10.1038/nphoton.2012.141},
	 
	number = {8},
	urldate = {2024-02-22},
	journal = {Nature Photon},
	author = {Ishikawa, Tetsuya and Aoyagi, Hideki and Asaka, Takao and Asano, Yoshihiro and Azumi, Noriyoshi and Bizen, Teruhiko and Ego, Hiroyasu and Fukami, Kenji and Fukui, Toru and Furukawa, Yukito and Goto, Shunji and Hanaki, Hirofumi and Hara, Toru and Hasegawa, Teruaki and Hatsui, Takaki and Higashiya, Atsushi and Hirono, Toko and Hosoda, Naoyasu and Ishii, Miho and Inagaki, Takahiro and Inubushi, Yuichi and Itoga, Toshiro and Joti, Yasumasa and Kago, Masahiro and Kameshima, Takashi and Kimura, Hiroaki and Kirihara, Yoichi and Kiyomichi, Akio and Kobayashi, Toshiaki and Kondo, Chikara and Kudo, Togo and Maesaka, Hirokazu and Maréchal, Xavier M. and Masuda, Takemasa and Matsubara, Shinichi and Matsumoto, Takahiro and Matsushita, Tomohiro and Matsui, Sakuo and Nagasono, Mitsuru and Nariyama, Nobuteru and Ohashi, Haruhiko and Ohata, Toru and Ohshima, Takashi and Ono, Shun and Otake, Yuji and Saji, Choji and Sakurai, Tatsuyuki and Sato, Takahiro and Sawada, Kei and Seike, Takamitsu and Shirasawa, Katsutoshi and Sugimoto, Takashi and Suzuki, Shinsuke and Takahashi, Sunao and Takebe, Hideki and Takeshita, Kunikazu and Tamasaku, Kenji and Tanaka, Hitoshi and Tanaka, Ryotaro and Tanaka, Takashi and Togashi, Tadashi and Togawa, Kazuaki and Tokuhisa, Atsushi and Tomizawa, Hiromitsu and Tono, Kensuke and Wu, Shukui and Yabashi, Makina and Yamaga, Mitsuhiro and Yamashita, Akihiro and Yanagida, Kenichi and Zhang, Chao and Shintake, Tsumoru and Kitamura, Hideo and Kumagai, Noritaka},
	month = aug,
	year = {2012},
	pages = {540--544},
}

@article{okuchi_ultrafast_2021,
	title = {Ultrafast olivine-ringwoodite transformation during shock compression},
	volume = {12},
	issn = {2041-1723},
	url = {https://www.nature.com/articles/s41467-021-24633-4},
	doi = {10.1038/s41467-021-24633-4},
	 
	number = {1},
	urldate = {2024-02-27},
	journal = {Nat Commun},
	author = {Okuchi, Takuo and Seto, Yusuke and Tomioka, Naotaka and Matsuoka, Takeshi and Albertazzi, Bruno and Hartley, Nicholas J. and Inubushi, Yuichi and Katagiri, Kento and Kodama, Ryosuke and Pikuz, Tatiana A. and Purevjav, Narangoo and Miyanishi, Kohei and Sato, Tomoko and Sekine, Toshimori and Sueda, Keiichi and Tanaka, Kazuo A. and Tange, Yoshinori and Togashi, Tadashi and Umeda, Yuhei and Yabuuchi, Toshinori and Yabashi, Makina and Ozaki, Norimasa},
	month = jul,
	year = {2021},
	pages = {4305},
}

@article{alexander_changes_2008,
	title = {Changes to the shock response of fused quartz due to glass modification},
	volume = {35},
	issn = {0734743X},
	url = {https://linkinghub.elsevier.com/retrieve/pii/S0734743X08001887},
	doi = {10.1016/j.ijimpeng.2008.07.019},
	 
	number = {12},
	urldate = {2024-02-27},
	journal = {International Journal of Impact Engineering},
	author = {Alexander, C.S. and Chhabildas, L.C. and Reinhart, W.D. and Templeton, D.W.},
	month = dec,
	year = {2008},
	pages = {1376--1385},
}

@article{schmelzer_crystallization_2016,
	title = {Crystallization of {Glass}: {What} {We} {Know}, {What} {We} {Need} to {Know}},
	volume = {7},
	issn = {2041-1286, 2041-1294},
	shorttitle = {Crystallization of {Glass}},
	url = {https://ceramics.onlinelibrary.wiley.com/doi/10.1111/ijag.12212},
	doi = {10.1111/ijag.12212},
	number = {3},
	urldate = {2024-03-07},
	journal = {Int J of Appl Glass Sci},
	author = {Schmelzer, Jürn W. P. and Fokin, Vladimir M. and Abyzov, Alexander S.},
	month = sep,
	year = {2016},
	pages = {253--261},
}

@article{carr_localized_2004,
	title = {Localized {Dynamics} during {Laser}-{Induced} {Damage} in {Optical} {Materials}},
	volume = {92},
	issn = {0031-9007, 1079-7114},
	url = {https://link.aps.org/doi/10.1103/PhysRevLett.92.087401},
	doi = {10.1103/PhysRevLett.92.087401},
	number = {8},
	urldate = {2024-03-07},
	journal = {Phys. Rev. Lett.},
	author = {Carr, C. W. and Radousky, H. B. and Rubenchik, A. M. and Feit, M. D. and Demos, S. G.},
	month = feb,
	year = {2004},
	pages = {087401},
}

@book{callister_fundamentals_2022,
	edition = {Sixth edition, international adaption},
	title = {Fundamentals of materials science and engineering: SI version},
	isbn = {978-1-119-82054-3},
	shorttitle = {Fundamentals of materials science and engineering},
	publisher = {Wiley},
	author = {Callister, William D. and Rethwisch, David G.},
	year = {2022},
}

@article{huang_two-stage_2003,
	title = {Two-{Stage} {Crystal}-{Growth} {Kinetics} {Observed} during {Hydrothermal} {Coarsening} of {Nanocrystalline} {ZnS}},
	volume = {3},
	issn = {1530-6984, 1530-6992},
	url = {https://pubs.acs.org/doi/10.1021/nl025836%2B},
	doi = {10.1021/nl025836+},
	 
	number = {3},
	urldate = {2024-04-09},
	journal = {Nano Lett.},
	author = {Huang, Feng and Zhang, Hengzhong and Banfield, Jillian F.},
	month = mar,
	year = {2003},
	pages = {373--378},
}

@article{ramis_multi_1988,
	title = {{MULTI} — {A} computer code for one-dimensional multigroup radiation hydrodynamics},
	volume = {49},
	issn = {00104655},
	url = {https://linkinghub.elsevier.com/retrieve/pii/0010465588900082},
	doi = {10.1016/0010-4655(88)90008-2},
	 
	number = {3},
	urldate = {2021-07-08},
	journal = {Computer Physics Communications},
	author = {Ramis, R. and Schmalz, R. and Meyer-Ter-Vehn, J.},
	month = jun,
	year = {1988},
	pages = {475--505},
}

@incollection{rice_compression_1958,
	title = {Compression of {Solids} by {Strong} {Shock} {Waves}},
	volume = {6},
	isbn = {978-0-12-607706-3},
	url = {https://linkinghub.elsevier.com/retrieve/pii/S0081194708607249},
	urldate = {2021-07-07},
	booktitle = {Solid {State} {Physics}},
	publisher = {Elsevier},
	author = {Rice, M.H. and McQueen, R.G. and Walsh, J.M.},
	year = {1958},
	doi = {10.1016/S0081-1947(08)60724-9},
	pages = {1--63},
}

@misc{wu_hugoniot_2024,
	title = {Hugoniot equation of state and sound velocity of {CaSiO3} glass under shock compression},
	copyright = {Creative Commons Attribution Non Commercial No Derivatives 4.0 International},
	url = {https://arxiv.org/abs/2412.13417},
	doi = {10.48550/ARXIV.2412.13417},
	urldate = {2025-02-25},
	publisher = {arXiv},
	author = {Wu, Ye and Zhang, Qing and Wang, Yishi and Hu, Yu and Li, Zehui and Li, Zining and Gao, Chang and Liu, Xun and Huang, Haijun and Fei, Yingwei},
	year = {2024},
	note = {Version Number: 1},
}

@article{tomioka_shockinduced_2000,
	title = {Shock‐induced transition of {NaAlSi}$_{\textrm{3}}$ {O}$_{\textrm{8}}$ feldspar into a hollandite structure in a {L6} chondrite},
	volume = {27},
	copyright = {http://onlinelibrary.wiley.com/termsAndConditions\#vor},
	issn = {0094-8276, 1944-8007},
	url = {https://agupubs.onlinelibrary.wiley.com/doi/10.1029/2000GL008513},
	doi = {10.1029/2000GL008513},
	number = {24},
	urldate = {2025-04-01},
	journal = {Geophysical Research Letters},
	author = {Tomioka, Naotaka and Mori, Hiroshi and Fujino, Kiyoshi},
	month = dec,
	year = {2000},
	pages = {3997--4000},
}

@article{schmelzer_crystallization_2015,
	title = {Crystallization of glass-forming liquids: {Maxima} of nucleation, growth, and overall crystallization rates},
	volume = {429},
	issn = {00223093},
	shorttitle = {Crystallization of glass-forming liquids},
	url = {https://linkinghub.elsevier.com/retrieve/pii/S0022309315301605},
	doi = {10.1016/j.jnoncrysol.2015.08.023},
	urldate = {2024-03-07},
	journal = {Journal of Non-Crystalline Solids},
	author = {Schmelzer, Jürn W.P. and Abyzov, Alexander S. and Fokin, Vladimir M. and Schick, Christoph and Zanotto, Edgar D.},
	month = dec,
	year = {2015},
	pages = {24--32},
}

@article{weinberg_location_1986,
	title = {On the location of the maximum homogeneous crystal nucleation temperature},
	volume = {83},
	issn = {00223093},
	url = {https://linkinghub.elsevier.com/retrieve/pii/0022309386900608},
	doi = {10.1016/0022-3093(86)90060-8},
	number = {1-2},
	urldate = {2025-12-18},
	journal = {Journal of Non-Crystalline Solids},
	author = {Weinberg, Michael C.},
	month = jun,
	year = {1986},
	pages = {98--113},
}

@article{warren_effect_1950,
	title = {The {Effect} of {Cold}-{Work} {Distortion} on {X}-{Ray} {Patterns}},
	volume = {21},
	issn = {0021-8979, 1089-7550},
	url = {https://pubs.aip.org/jap/article/21/6/595/521146/The-Effect-of-Cold-Work-Distortion-on-X-Ray},
	doi = {10.1063/1.1699713},
	number = {6},
	urldate = {2026-01-09},
	journal = {Journal of Applied Physics},
	author = {Warren, B. E. and Averbach, B. L.},
	month = jun,
	year = {1950},
	pages = {595--599},
}

@article{noguchi_high-temperature_2013,
	title = {High-temperature compression experiments of {CaSiO3} perovskite to lowermost mantle conditions and its thermal equation of state},
	volume = {40},
	issn = {0342-1791, 1432-2021},
	url = {http://link.springer.com/10.1007/s00269-012-0549-1},
	doi = {10.1007/s00269-012-0549-1},
	number = {1},
	urldate = {2022-08-09},
	journal = {Physics and Chemistry of Minerals},
	author = {Noguchi, Masanao and Komabayashi, Tetsuya and Hirose, Kei and Ohishi, Yasuo},
	month = jan,
	year = {2013},
	pages = {81--91},
}

@article{immoor_weak_2022,
	title = {Weak cubic {CaSiO3} perovskite in the {Earth}’s mantle},
	volume = {603},
	issn = {0028-0836, 1476-4687},
	url = {https://www.nature.com/articles/s41586-021-04378-2},
	doi = {10.1038/s41586-021-04378-2},
	number = {7900},
	urldate = {2023-06-27},
	journal = {Nature},
	author = {Immoor, J. and Miyagi, L. and Liermann, H.-P. and Speziale, S. and Schulze, K. and Buchen, J. and Kurnosov, A. and Marquardt, H.},
	month = mar,
	year = {2022},
	pages = {276--279},
}

@article{stixrude_phase_2007,
	title = {Phase stability and shear softening in {Ca} {Si} {O} 3 perovskite at high pressure},
	volume = {75},
	issn = {1098-0121, 1550-235X},
	url = {https://link.aps.org/doi/10.1103/PhysRevB.75.024108},
	doi = {10.1103/PhysRevB.75.024108},
	number = {2},
	urldate = {2024-03-13},
	journal = {Physical Review B},
	author = {Stixrude, Lars and Lithgow-Bertelloni, C. and Kiefer, B. and Fumagalli, P.},
	month = jan,
	year = {2007},
	pages = {024108},
}

@article{kawai_pvt_2014,
	title = {\textit{{P}‐{V}‐{T}} equation of state of cubic {CaSiO} $_{\textrm{3}}$ perovskite from first‐principles computation},
	volume = {119},
	copyright = {http://onlinelibrary.wiley.com/termsAndConditions\#vor},
	issn = {2169-9313, 2169-9356},
	url = {https://agupubs.onlinelibrary.wiley.com/doi/10.1002/2013JB010905},
	doi = {10.1002/2013JB010905},
	number = {4},
	urldate = {2024-05-03},
	journal = {Journal of Geophysical Research: Solid Earth},
	author = {Kawai, Kenji and Tsuchiya, Taku},
	month = apr,
	year = {2014},
	pages = {2801--2809},
}

@article{sun_high-pressure_2022,
	title = {High-pressure experimental study of tetragonal {CaSiO3}-perovskite to 200 {GPa}},
	volume = {107},
	issn = {0003-004X, 1945-3027},
	url = {https://pubs.geoscienceworld.org/ammin/article/107/1/110/610330/High-pressure-experimental-study-of-tetragonal},
	doi = {10.2138/am-2021-7913},
	number = {1},
	urldate = {2025-01-28},
	journal = {American Mineralogist},
	author = {Sun, Ningyu and Bian, Hui and Zhang, Youyue and Lin, Jung-Fu and Prakapenka, Vitali B. and Mao, Zhu},
	month = jan,
	year = {2022},
	pages = {110--115},
}

@book{christian_theory_2002,
	title = {The {Theory} of {Transformations} in {Metals} and {Alloys}},
	isbn = {978-0-08-044019-4},
	url = {https://linkinghub.elsevier.com/retrieve/pii/B9780080440194X50004},
	urldate = {2026-01-21},
	publisher = {Elsevier},
	author = {Christian, J.W.},
	year = {2002},
	doi = {10.1016/B978-0-08-044019-4.X5000-4},
}

@article{lifshitz_kinetics_1961,
	title = {The kinetics of precipitation from supersaturated solid solutions},
	volume = {19},
	copyright = {https://www.elsevier.com/tdm/userlicense/1.0/},
	issn = {00223697},
	url = {https://linkinghub.elsevier.com/retrieve/pii/0022369761900543},
	doi = {10.1016/0022-3697(61)90054-3},
	number = {1-2},
	urldate = {2026-02-03},
	journal = {Journal of Physics and Chemistry of Solids},
	author = {Lifshitz, I.M. and Slyozov, V.V.},
	month = apr,
	year = {1961},
	pages = {35--50},
}

\end{document}